\documentstyle[prabib, aps]{revtex}
\begin{document}
\draft
\title{Kolmogorov Algorithmic Complexity and its Probability \\
Interpretation in Quantum Gravity}
\author{V.D.Dzhunushaliev
\thanks{E-mail address: dzhun@freenet.bishkek.su}}
\address {Department of the Theoretical Physics \\
          Kyrgyz State National University, Bishkek, 720024}

\maketitle
\vspace{30mm}
\centerline{\bf \large Short Title}
\centerline{\bf \large Kolmogorov Algorithmic Complexity ...}
\vspace{30mm}

\newpage
\begin{abstract}
The quantum gravity has great difficulties with application of the 
probability notion. In given article this problem is analyzed according 
to algorithmic viewpoint. According to A.N. Kolmogorov, the probability 
notion can be connected with algorithmic complexity of given object. 
The paper proposes an interpretation of quantum gravity, according to 
which an appearance of something corresponds to its Kolmogorov's 
algorithmic complexity. By this viewpoint the following questions are 
considered: the  quantum transition with supplementary coordinates 
splitting off, the algorithmic complexity of the  Schwarzschild black hole 
is  estimated, the redefinition of the Feynman path  integral, the quantum  
birth  of  the Euclidean Universe  with  the following changing of the 
metric signature. 
\end{abstract}

\newpage

\section{Introduction}

\par
The spacetime with gravity and matter fields contains the notions of 
topological space, geometrical structures (metric + connection), signature 
of metric and so on. Some above - mentioned structures (for example, a 
topology on some space) cannot be the solutions of any differential 
equations. The other structures are the solutions of corresponding 
differential equations, for example, the metric defines from gravity 
equations, the gauge fields (connection) from Yang - Mills equations. 
Hence, it is necessary to distinguish the two different regions in quantum 
gravity:
\begin{enumerate}
\item
In the first region we have to do with questions connected with appearance 
of one or another nondifferentiable structures of spacetime until 
the appearance of some kind of differentiable structures on this 
topological space.
\item
In the second region we quantize the differentiable structures: 
gravity (metric) and gauge fields (connection).
\end{enumerate}
\par
It is clearly that the quantization in the first case cannot be carried out 
as in the second case. Intuitively, it seems that the probability of 
appearance one or another nondifferentiable structures from the first region 
can be connected with its complexity. That is, the more complicated 
structures have to appears with the smaller probability and conversely.
\par
In  60s  A.N. Kolmogorov investigated   the probability  notion  from 
the algorithmic  viewpoint.   The classical probability  definition  is  
connected  with calculation of a ratio of the number of ways in which the 
trial can succeed to the total number of ways in which the trial can result.
Kolmogorov  had investigated  the   notion   of probability from the 
algorithmic viewpoint. 
He  wrote  in\cite{kol}:
\begin{enumerate}
\item
the basic notions of the information theory must and  may
be defined without the application of probability theory, so that the 
``entropy'' and ``number of information'' notions are  applicable
to the \underline{individual objects};
\item
the notions of the information theory are introduced  by  this
manner can form the basis  of a new  conception of the 
chance, corresponding to natural idea that the chance is the absence 
of laws.
\end{enumerate}
\par
For  our  quantum  gravitational  purpose this Kolmogorov idea is very
important and the probability notion 
is adaptable to single object  (one or another 
nondifferentiable structures on Universe).  From 
Kolmogorov's viewpoint {\em ``chance'' = ``complexity''}, and  therefore  
the probability in quantum gravity (QG) can be  connected  with
the complexity of a given structure. Now we  shall  give  
the exact definition of the algorithmic  complexity  ($AC$) us
Kolmogorov did \cite{kol}:
{\em
\par
The algorithmic complexity $K(x\mid y)$ of the  object $x$  by 
given object $y$ is the minimal length of the ``program'' $P$
that is written as a sequence of  the  zeros  and  unities
which allows as to construct $x$ having $y$:
\begin{equation}
K(x\mid y) = \min_{A( P,y)=x} l(P)
\label{11}
\end{equation}
where $l(P)$ is length of the  program $P$; $A(P,y)$  is  the
algorithm  calculating  object $x$, using  the  program $P$,
when the object $y$ is given.
}
\par
According to this definition, it can  assume that  some 
Universe structures are appeared (in QG) 
with the probability depending from its $AC$ by the following manner:
\begin{equation}
P[g] \approx  P_{0} e^{ -K[g]},
\label{12}
\end{equation}
where $P_{0}$ is the normalized constant; $K[g]$ is the $AC$ of given 
object. For example, it can  assume that  the  Universe
birth probability from ``Nothing''  by Eq.(\ref{12}) is defined,
where $K[g]$ is $AC$ of Universe.  Furthermore, we  shall  assume
that the $AC$ of given object can be spontaneously changed. In this 
case the situation  is analogous to the spontaneous electron transition 
between  two energy levels in atom: electron that was on the  exited-state
level $E_{1}$ (quantum gravitational object with the $AC$ $K_{1}$) falls 
in  the result of the spontaneous quantum  transition  on  the
ground-state  level $E_{2}$ (appear  the  quantum  gravitational
object with lesser $AC$ $K_{2}< K_{1}$). It is highly probable that these
processes of the quantum birth of Universe 
and the quantum transition with the $AC$ changing happen in  Plank
region.  The  splitting  off of  the  supplementary
dimension; the changing of the metric signature, topology  and
so on can be such quantum transition with the $AC$ changing.
\par
Thus, the  basic  physical  idea  given  here is:
{\em
\par
In the Plank region can happen both  the  spontaneous
quantum birth of the quantum gravitational object having 
the fixed $AC$  and  the  spontaneous  quantum  changing of it
properties with corresponding changing the $AC$.}
\par
Certainly it is  necessary  to sew all physical quantities  
by $AC$ changing of this object. Let's consider some questions of QG and 
attempt to  connect the $AC$  notion  with  the  corresponding  quantum 
probability.

\section
{The quantum transitions from multidimensional to 4D Universe 
with supplementary coordinates splitting off.}

In this section the quantum transitions model is offered
with splitting off of the supplementary  dimensions  ($SD$)
in gravity during Universe evolution.
\par
Thus, we suppose that the regions can exist in the spacetime
in which $SD$ turn on (off). In algorithmic viewpoint this
means that we  must  describe the  whole  spacetime  by  not  one
algorithm (one field equations system) but more than one: 
each region has its nature laws (system of  the  field  equation).
Naturally, all field variables must be sewed on the boundaries of 
this regions. We consider the model example, in which the $4D$
Universe arises in the result of quantum spontaneous
transition from the empty multidimensional Kazner's  Universe
after splitting off $SD$ \cite{dzh2}.

\subsection
{7-dimensional cosmological solution.}

Let's consider the empty $7D$ spacetime with  the  metrics
in the following form:
\begin{equation}
ds^{2} = dt^{2} - a^{2}(t) dl^{2}_{1} - b^{2}(t) dl^{2}_{2},
\label{51}
\end{equation}
where $dl^{2}_{1,2}$ is metrics on $S^{3}$ (3D sphere), $L^{3}$ (3D  
Lobachewsky space) or $E^{3}$ (3D  Euclidean  space).  The  Einstein's  
vacuum equations have the following form:
\begin{mathletters}
\begin{eqnarray}
\frac {\ddot a}{a} = -\frac {\dot a ^{2}}{a^{2}} +
                     \frac {\dot b ^{2}}{b^{2}} -
                     \frac {k_{1}}{a^{2}} +
                     \frac {k_{2}}{b^{2}},
\label{521}\\
\frac {\ddot b}{b} = -\frac {\dot b ^{2}}{b^{2}} +
                     \frac {\dot a ^{2}}{a^{2}} +
                     \frac {k_{1}}{a^{2}} -
                     \frac {k_{2}}{b^{2}},
\label{522}\\
3\frac{\dot a \dot b}{ab} + \frac {\dot a ^{2}}{a^{2}} +
                           \frac {\dot b ^{2}}{b^{2}} +
                           \frac {k_{1}}{a^{2}} +
                           \frac {k_{2}}{b^{2}} = 0,
\label{523}
\end{eqnarray}
\end{mathletters}
where  $(\dot{\phantom  x})$   means   the   time
derivative of $t$; $k_{1,2}~=~\pm 1,~0$
respectively for $S^{3}$, $L^{3}$ and $E^{3}$. 
We consider the simplest case for which it  is  possible
to find  analytical solution: $k_{1,2}=0$.  In  this  case  system
(\ref{521}-\ref{523}) has the following exact solution:
\begin{mathletters}
\begin{eqnarray}
a = a_{0} \left( - \frac{t}{a_{1}} \right) ^{\alpha};
\qquad
b = b_{0} \left( - \frac{t}{b_{1}} \right) ^{\beta}
\label{531}\\
\alpha = \frac{1 - \sqrt{5}}{6};
\qquad
\beta = \frac{1 + \sqrt{5}}{6}
\label{532}
\end{eqnarray}
\end{mathletters}
where $a_{0,1}$ and $b_{0,1}$ are the constants that should be defined
later by discussion of question of splitting off $SD$; $t<0$.

\subsection{4-dimensional Universe.}

First, we  define  the  form  of  the  Einstein's
equations in $4D$ spacetime after $SD$ splitting off.  We  should
start from the multidimensional Einstein's  Lagrangian  
describing the dynamics before the $SD$ splitting off:
\begin{equation}
- L_{7D} = \sqrt{G} R_{7D} = \sqrt{G}\left[ R_{4D} +(cross\; term) + R_{SD}
  \right],
\label{54}
\end{equation}
where $G$ is determinant of the $7D$ metrical tensor, $R_{4D}$  is $4D$
scalar curvature of the  Einstein's  dimensions, the  second
term on the right part of Eq.(\ref{54}) describes the  metric dependence 
on $SD$ from the Einstein's coordinates, $R_{SD}$  is
the scalar curvature of the $SD$  space. The metric on the 
supplementary coordinates $\gamma _{ab}$ does not 
vary after the $SD$ splitting off, that  is $\gamma _{ab}$~=~const. 
This leads to the fact that  now  (after $SD$  splitting  off) $4D$
Lagrangian has the following form:
\begin{equation}
- L_{4D} = \sqrt{-g}\left ( R_{4D} -\Lambda \right ),
\label{55}
\end{equation}
where $g$ is determinant $4D$  metrics; 
$\Lambda = R_{SD} = 6k_{2}/r^{2}_{0}$ = const
(it is suggested that the $SD$ is $S^{3}$, $L^{3}$ or $E^{3}$), $r_{0}$ is 
some characteristic length on the $SD$ (later it will be showed
that $r_{0}~=~b_{0})$. By this manner after the $SD$ splitting off,  we
have $\Lambda $  -  term that  has  a  physical  meaning  of  scalar
curvature of supplementary coordinates. $4D$ metric is being searched in 
the form:
\begin{equation}
ds^{2} = dt^{2} - a^{2}(t) dl^{2}
\label{56}
\end{equation}
where element $dl^{2}$ is similar to $dl^{2}_1$ in Eq.(\ref{51}).
\par
The Einstein's equations for this metrics are:
\begin{mathletters}
\begin{eqnarray}
2\frac{\ddot a}{a} + \frac{\dot a ^{2}}{a^{2}} +
 \frac{k_{1}}{a^{2}} + 3\frac{k_{2}}{r^{2}_{0}} = 0,
\label{571}\\
\frac{\dot a^{2}}{a^{2}} + \frac{k_{1}}{a^{2}} +
\frac{k_{2}}{r^{2}_{0}} = 0,
\label{572}
\end{eqnarray}
\end{mathletters}
where $k_{2}/r^{2}_{0}$ is the scalar  curvature
 of  the $SD$  after  its
splitting off. For $k_{1,2} = 0$ we have the following solution:
\begin{equation}
a = a'_{0} = const.
\label{58}
\end{equation}
This space is the Minkowski spacetime.
\par
According to Kolmogorov's definition both multidimensional
and $4D$ Universes considered  above  have the various $AC$.
Obviously, that the $AC$ of the multidimensional  Universe  described
by system (\ref{521}-\ref{523}) is more than the $AC$ of $4D$ Universe, 
because the equations number in the system (\ref{521}-\ref{523}) is 
more than that in system (\ref{571}-\ref{572}).
\par
Let's consider the following model describing  the $4D$
Universe birth: At the beginning the Universe  is  the  empty
multidimensional Kazner's Universe evolving according to  Eq.
(\ref{531}-\ref{532}) to the singularity $t = 0$.  Near  the  
singularity $a(t)\to \infty$ and $b(t)\to 0$. Therefore, the terms
$(\dot a /a)^{2}$, $(\dot b /b)^{2}$
$(\ddot a /a)^{2}$ , $(\ddot b /b)^{2}$,
are comparable with Plank curvature and make
a key contribution to the scalar curvature $R_{7D}$. It  leads  to
that the quantities $\dot a$   and  $\dot b$   have  the
essential  quantum fluctuations.
\par
When  the quantities  $a/\dot a$ become
comparable with the Plank length, the  quantum  transition 
happens  from  multidimensional  equations  (\ref{521}-\ref{523}) to  
the $4D$ equations (\ref{571}-\ref{572}). This process occurs due to 
the fact that the $AC$ of the $4D$ Universe is less  than the $AC$  of  the  
multidimensional Universe.
\par
Let's choose the next value for constants $a_{0,1}$ and $b_{0,1}$
in the  Eq.(\ref{531}-\ref{532}): 
$a_{0}\gg b_{0}\gg l_{pl}$, $a_{1} = b_{1} = l_{pl}$.
Then  for $t~\approx ~t_{pl}$:
\begin{mathletters}
\begin{eqnarray}
a(t_{pl}) \approx  a_{0};\qquad b(t_{pl}) \approx  b_{0},
\label{591}\\
\frac{\dot a ^{2}}{a^{2}}\approx \frac{\dot b ^{2}}{b^{2}}\approx
\frac{\ddot a}{a}\approx \frac{\ddot b}{b}\approx \frac{1}{l^{2}_{Pl}}.
\label{592}
\end{eqnarray}
\end{mathletters}
After this transition the following facts take place:
\begin{enumerate}
\item
The linear $4D$ scale $a_{0}$ and $b_{0}$ (respectively
in Einstein's
and supplementary dimension) remain in the  classical  region
and they become  equal  to  the  respective  multidimensional
values before the $SD$ splitting off;
\item
$4D$ Universe evolves according to the Einstein's equations
(\ref{571}-\ref{572});
\item
The $SD$ metrics  becomes  non-dynamical  variable  and  doesn't
change with time.
\end{enumerate}
\par
Let's estimate the probability $P$  of  given  transition
from multidimensional Universe to the $4D$ Universe as:
\begin{equation}
P_{multiD\to 4D} = \frac{e^{-K_2}}
             {e^{-K_1} + e^{-K_2}},
\label{510}
\end{equation}
where $K_{1,2}$ in (\ref{510}) are respectively the $AC$ of
multidimensional and $4D$ Universes described in Eq's.(\ref{521}-\ref{523})
and (\ref{571}-\ref{572}). According to the $AC$ definition $K_{1,2}$ is 
calculated as a program length  in  bits  for  some  universal  machine  
(for example, for the Turing machine). It is known  that  the
program describing even such simple arithmetical operation as
addition in the Turing machine is very  large.  As  a  result, 
$K_{1}\gg K_{2}$. Then Eq.(\ref{510}) can be  approximately  estimated  as
follows:
\begin{equation}
P_{multiD\to 4D}\approx 1 - e^{-\left[ K_1 - K_2 
            \right ]} \approx 1.
\label{511}
\end{equation}

\section{The electrical charge model with splitting off\\
	 the supplementary coordinates}

The aim of this section is the construction of a composite wormhole:
it is the solution of 5D Einstein's equations (Kaluza - Klein's equation)
under event horizon ($EH$), and the Reissner - Nordstr\"om's solution
outside $EH$. The sewing the 5D Kaluza - Klein's 
metric with 4D Einstein's metric + electrical field is performed on $EH$, 
that is, splitting off the supplementary coordinates happens.

\subsection{5D Wormhole in Kaluza - Klein's theory}

At first we remind the results achieved in Ref.\cite{dzh1}.  
$5D$ metric has the following wormhole-like view:
\begin{mathletters}
\begin{equation}
ds^2 = e^{2\nu (r)}dt^2 - e^{2\psi (r)} \left (d\chi - \omega (r)dt
\right )^2 - dr^2 - a^2(r)\left (d\theta ^2 + \sin ^2\theta d\varphi ^2
\right ),
\label{11a}
\end{equation}
\end{mathletters}
where $\chi$ is 5 supplementary coordinate; $r, \theta ,\varphi$ 
are 3D polar coordinates; $t$ is time. Corresponding $5D$ Einstein's  
equations have the following solution:
\begin{mathletters}
\begin{eqnarray}
a^2 & = & r^2_0 + r^2,
\label{12a}\\
e^{-2\psi}= e^{2\nu} & = & {{2r_0}\over q} {{r^2_0 + r^2}
\over{r^2_0 - r^2}},
\label{12b}\\
\omega &= & {{4r^2_0}\over q} {r\over{r^2_0 - r^2}},
\label{12c}
\end{eqnarray}
\end{mathletters}
where $r_0$ is a throat of this wormhole; $q$ is a 5D 
''electrical'' charge. It is easy to see that the time component of 
metrical tensor $G_{tt} (r=\pm r_0)=0$. This indicates that there is the 
null surface, as on its $ds^2=0$. The sewing of the $5D$ and $4D$ physical 
quantities happens in the following manner:
\begin{mathletters}
\begin{eqnarray}
e^{2\nu _0} - \omega ^2_0 e^{-2\nu _0} = G_{tt}\left 
(\pm r_0\right ) = g_{tt}\left (r_+\right ) = 0,
\label{13a}\\
r^2_0 = G_{\theta\theta}(\pm r_0) = g_{\theta\theta}(r_+) = r^2_+,
\label{13b}
\end{eqnarray}
\end{mathletters}
where $G$ and $g$ are $5D$ and $4D$ metrical tensors, respectively. 
$r_+ = m + \sqrt{m^2 + Q^2}$ is the $EH$ for Reissner - Nordstr\"om's
solution ($m$ and $Q$ are mass and charge of the Reissner - Nordstr\"om's
black hole). The quantities marked by $(0)$ sign are taken by $r=\pm r_0$.

\subsection{The sewing Kaluza - Klein's wormhole with 
Reissner - Nordstr\"om's solution}

In this section we show that the Reissner - Nordstr\"om's solution
really can sew on $EH$ with above received 5D wormhole. 
To do this, the 5D Kaluza - Klein's metrical tensor is necessary 
to sew with (4D metric + Maxwell's electrical field in Reissner - 
Nordstr\"om's solution).
\par
To sew $G_{\chi t}$ and 4D electrical field we consider 5D 
$R_{\chi t}=0$ and Maxwell's equations:
\begin{mathletters}
\begin{eqnarray}
\left [a^2\left (\omega 'e^{-4\nu}\right )\right ]' = 0,
\label{23}\\
\left (r^2E_r\right )' = 0,
\label{24}
\end{eqnarray}
\end{mathletters}
here $E_r$ is 4D electrical field.
\par
These two equations are practically the Gauss law and they indicate that
some quantity multiplied by area is conserved. In 4D case this quantity 
is 4D Maxwell's electrical field and from this follows that the 
electrical charge is 
conserved. Thus, naturally we must join 4D electrical 
Reissner - Nordstr\"om's 
field $E_{RN} = Q/r^2_+$ with ``electrical'' Kaluza - Klein's field 
$E_{KK} = \omega 'e^{-4\nu}$ on $EH$:
\begin{equation}
\omega _0'e^{-4\nu _0} = {q\over{2r^2_0}} = E_{KK} = 
E_{RN} = {Q\over{r^2_+}}.
\label{25}
\end{equation}
In this case the probability of splitting off the $SD$ is calculated 
similar to Eq.(\ref{510}) definition.

\section
{The algorithmic complexity of the \\ Schwarzschild black hole.}

Beckenstein \cite{bek1}, \cite{bek2} and Hawking \cite{haw} have showed 
that the  black hole has entropy  connected  with  the  existence  of  event
horizon. The  entropy  notion  is  usually  employed  to  the
statistical  systems  consisted of a great number  of 
particles. But in given case this notion  is  used  with  the
individual object as it was proposed by Kolmogorov.
\par
According to the definition  (\ref{11}), the $AC$ 
is defined as  the  smallest  algorithm  describing  a  given
gravity field ($GF$). The metrical tensor is  the  solution  of
the gravity equations (Einstein's equations, $R^{2}$ -  theory  or
the  gauge  gravity  equations).  Therefore, the   algorithm
describing  given  metrics  is  the   corresponding   gravity
equations restoring the $GF$ at  the  all  spacetime  from  the
initial  and(or)  boundary  conditions.  In  this   case   the
algorithm length describing $GF$ is essentially reduced.
\par
Now we shall calculate $AC$ for the  Schwarzschild  black
hole. For this purpose we shall write the metric in the following 
form:
\begin{equation}
ds^{2} = dt^{2} - e^{\lambda (t,R)}dR^{2} - r^{2}(t,R) 
\left( d\theta ^{2} + \sin ^{2} \theta d\phi ^{2} \right),
\label{41}
\end{equation}
where $t$ is time, $R$ is radius, $\theta$ and $\phi$ are  polar  angles.
In this case Einstein's equations are:
\begin{mathletters}
\begin{eqnarray}
-e^{-\lambda}r'^2 + 2r\ddot r + \dot r ^2 + 1 & = & 0,
\label{42a}\\
-\frac{e^{-\lambda}}{r}\left (2r'' - r'\lambda ' \right ) + 
\frac{\dot r \dot \lambda}{t} + \ddot \lambda + 
\frac{\dot \lambda ^2}{2} + 
\frac{2\ddot r}{r} & = & 0,
\label{42b}\\
-\frac{e{-\lambda}}{r^{2}}
 \left(2rr'' + {r'} ^{2} -rr'\lambda '\right) +
 \frac{1}{r^{2}}\left( r\dot a \dot \lambda + {\dot a} ^{2} + 
1 \right) & = & 0,
\label{42c}\\
2\dot r' - \dot \lambda r' & = & 0,
\label{43d}
\end{eqnarray}
\end{mathletters}
where $(')$ and $(\dot{\phantom x})$ mean respectively the derivations
on $t$ and $r$. We give the wormhole $t=0$ as a Caushy  hypersurface.  
The assignment of initial data on this hypersurface define $GF$ 
on the whole  Schwarzschild spacetime. But ``quantity'' of the
initial data can be essentially reduced. Actually, we  examine
the Eq.(\ref{42c}) on  the  wormhole $t=0$.  Here,  first  time
derivative of all components of the metrical tensor is equal
to zero. Therefore the initial data must satisfy to equation:
\begin{equation}
2rr''  + {r'} ^{2} - rr' \lambda '  - e^{\lambda } = 0.
\label{43}
\end{equation}
To solve Eq. (\ref{43}) on surface $t=0$  it  is  necessary  to
define the boundary conditions. They may be taken on throat of wormhole 
in the following form:
\begin{equation}
r' (R=0,t=0) = 0,\qquad r(R=0,t=0) = r_{g},
\label{44}
\end{equation}
where $r_{g}$ is radius at the event horizon. It means that the $GF$
on the whole Schwarzschild spacetime is defined by the value
of $G_{\theta\theta}$ component of metrical tensor at the origin.
Therefore, the $AC$ for the Schwarzschild metrics  is
defined by the following expression:
\begin{equation}
K \approx  L\left[ \left (\frac{r_{g}}{r_{Pl}} \right) ^2 \right ]
+ L_{Einstein}, 
\label{46}
\end{equation}
where $L\left [({r^{2}/r^{2}_{Pl}})\right ]$ is  program length of  the  
definition  of dimensionless number $r^{2}_{g}/r^{2}_{Pl}$ made up 
for the  some  universal machine; $L_{Einstein}$ is  program  length  of  
the  solution  of Einstein's differential equations using some universal 
machine, for example, the Turing machine.

\section{Algorithmic complexity and the path integral}

In this section we will obtain well defined positive  functional  on  the
metric space that can be important for QG. On this basis 
we define the path integral in $QG$ by the following  manner  \cite{dzh1}:
The action functional in the path integral we replace with $AC$
for a given  metric,  which is the positive defined functional of the 
metric:
\begin{equation}
\int  D[g] e^{-i(I[g] + \int g_{\mu\nu}J^{\mu\nu}dx)} \rightarrow  
\int  D[g] e^{-i(K[g] + \int g_{\mu\nu}J^{\mu\nu}dx)} = 
e^{iZ[J^{\mu\nu}]},
\label{21a}
\end{equation}
where $g_{\mu\nu}$ is a metric; $K[g]$ is the $AC$ of given metric $g$; 
$Z[J]$ is a generating functional for QG.
\par
It  is  sufficiently  obvious  that  the  most   complicated 
gravitational fields are the metrics satisfying none 
field equations. According  to  Kolmogorov, they  are  random
fields: in consequence of absence of the  algorithm  connecting  the
metric value in the neighbouring points. The metrics, which are  
the solutions of some gravity  equations  (Einstein's, $R^{2}$  -
theory, Euclidean, multidimensional and so on) have a much lesser 
$AC$ in comparison with random metrics as  they  are  calculated
from  initial  and/or  boundary  conditions. The gravitational instantons 
are  the  simplest  gravity  objects: they are symmetrical  spaces  
with the  corresponding   metrics possessing the some symmetry group. 
It is necessary to  mark that the instantons can be defined not by field 
equations but by its  topology  charges that strongly simplifies its 
algorithmic definitions.
\par
Thus, for a first approximation the path integral  in QG
is defined by sum on all  gravity  instantons.  In  the  next
approximation the contributions from the  metrics  which  are
the solutions of the Einstein's  equations, $R^{2}$  -  theories,
multidimensional  theories  and  etc. are appeared. The
contribution from complicated metrics (in   above-mentioned
Kolmogorov's meaning) appears when the quantum field effects of 
gravitational field are takes into  account. In the QG  based  on  the
integral (\ref{21a}) the Universe can contain a few regions 
with various gravity equations.
\par
The effects  connected  with  topology  changing (spacetime  foam)
in the sum (\ref{21a}) are taken automatically into  account, 
because it  is  quite  evidently  that the  topological  different
spaces have the different $AC$.
\par
According to this idea, we could suppose that the  changing 
of the Universe $AC$ is possible during the process
of  its  evolution. The Universe is born  from  ``Nothing'' with 
certain $AC$ (it can be multidimensional and/or Euclidean)  
and then in own evolution process it can happen the processes 
with changing of its $AC$: the supplementary coordinates splitting off;
the metric signature changing; the simultaneous suppressing
term like $R^{2}$ or $R^{a}_{bij} R_{a}^{bij}$ (arising 
in the gauge theories of  gravity) in Lagrangian and so on.
J. Wheeler \cite{wh} used the similar idea discussing the 
question of appearance of our Universe  from  ``Pregeometry''.

\section
{The quantum birth of the  Euclidean  Universe
with the following changing of the metric signature.}

Every physical object in the nature may be described  by
some algorithm.  For  example,  the  Universe  in  Einstein's
theory is  characterized  by  algorithm  defining  all
notions laying at the base of  the  Universe.  In  this  case 
the topological  space, its  geometrical  structures   (metrics,
connection), Einstein's equations and so on must be described. Such
algorithm should be realized with some universal machine (for
example, with Turing machine). As it  was  mentioned  above, 
the length of such minimal algorithm is called the $AC$ of  given
object and is defined by (\ref{11}).
\par
The Euclidean instantons have the simplest  construction among  all smooth 
manifolds with metrics and  connection.  If the algorithm is describing 
differential  manifold  and geometrical structure on spacetime to fix, then 
it is remaining to define the  algorithm   determining the  concrete  metric 
and connection only. In  any geometrical gravity theory such algorithm  will 
be  field gravitational equations. In this case it is possible to build the 
following hierarchy in accordance with increasing  complexity: Euclidean  
instantons, the  spacetimes  satisfying to any gravity equations  (for  
example  Einstein's  equations)  and other random spaces. We remind that 
the gravity equation is some algorithm  computing  the  metric  and  
connection  from  the initial and(or) boundary conditions.
\par
Thus, we  concluded that  some Euclidean instanton 
from the gravity vacuum with the greatest probability in consequence 
of  quantum fluctuations is  appeared. We suppose that it is born from 
the  vacuum  in  the cut form. For  example,  de'Sitters  instanton $S^{4}$ 
($4D$sphere) can appear without neighbourhood of some 
point. This cut instanton has a boundary. Let's suppose  that the
spontaneous changing of the metrics signature  takes  place  on
the boundary and the pseudo-Riemannian spacetime is  appeared
that begins its evolution from boundary according to one
or another gravity equations. This is well known Hawking's idea on the 
Universe birth from Euclidean region from the algorithmical viewpoint.
\par
Now let's define the $AC$ of the instanton. The instanton   is
described   by   the   self-duality   equations   which   are
differential equations of the first order,  while  the  field
gravity  equations  describing  spacetime  are the differential 
equations of the second  order. Consequently, the instantons  are   
simpler objects than another spacetimes.  Let's notice 
that the instanton can be described in still  simpler  manner.  Indeed, 
the instanton is completely defined by its topological number
- Pontragin's index. Therefore the $AC$ of  instanton  is
equal to the length of the algorithm describing  the  integer
$n$~=~Pontragin's index. Thus:
\begin{equation}
K[instanton] = (the\; record\; length\; of\; integer\; n) + K_{0}
\label{31}
\end{equation}
where $K_{0}$ is the length of the  algorithm  describing  all
other  notions:  the  topological  space, the   geometrical
structure on it and so on.
\par
The  algorithm  length  describing  integer $n$  can be estimated  in  the  
following  way:  The  integer $n$ can be written as $n$  ``stick-unity''.  
Then, the considered algorithm consists of elementary steps, where each  of 
them calculates one ``stick-unity'' from  $n$. Then the length of this 
algorithm is approximately  equal  to the examined integer $n$.
\par
Let's consider the quantity $exp (- instanton\; action)$ 
connected  with  the  probability   of corresponding tunnel
transition.  It  is well known  that  the  instanton  action   is
proportional to the integer $n$. Thus two integers ($AC$ and  the
instanton  action)  coincide  up  to  a  constant.  This is a 
supplementary argument for support of the (\ref{21a}) expression.
\par
If our algorithm could  be  broken  into  several  steps
describing  metrics $g$  step  by  step,  then  the   complete
probability  is  equal  to  the  product  of  the  applicable
probability for each separate step. For example,  
the  whole  algorithm describing the Universe includes
consistently the algorithms describing the topological  space, 
the  differentiable  structure on it,  the  geometrical   structure
(metrics and connections), the field gravity  equations  and so on.
For our purposes we single out the  separate  factor, describing 
the algorithm computating the metrics
and connection according to the field equations. The remained
factor is one for all Universes and it can be  inserted  into
the normalization constant.
\par
Let's propose that the changing of the  metrics  signature
happens with the following probability:
\begin{equation}
P_{change} = \frac{e^{-K[Mink]}}
             {e^{-K[Eucl]} + e^{-K[Mink]}}
\label{33}
\end{equation}
where $K[Eucl]$ and $K[Mink]$ are accordingly $AC$ of  the  gravity
equations in the Euclidean and pseudo-Riemannian spaces. It is necessary to 
note that this process is essentially quantum phenomenon rather than 
classical.

\section{Acknowledgments}

This research was supported by ISF Grant MYT000.


\begin{references}

\bibitem{kol}
A.N. Kolmogorov, Information theory and  algorithm  theory:
Logical foundation of the information  theory.  Combinatorial
foundation of the  information  theory  and  the  probability
calculus. (M.: Nauka, 1987).
\bibitem{dzh2}
V.D.Dzhunushaliev, Izv. vuzov, ser. Fizika, N9, 55(1994) (In Russian).
\bibitem{dzh}
V.D.Dzhunushaliev, Izv. vuzov, ser. Fizika, N6, 78(1993) (In Russian).
\bibitem{bek1}
J.D.Bekenstein, Phys.Rev.{\bf D7}, 2333(1973).
\bibitem{bek2}
J.D.Bekenstein, Phys.Rev.{\bf D9}, 3292(1974).
\bibitem{haw}
S.W.Hawking, Nature, {\bf 238}, 30(1974).
\bibitem{dzh1}
V.D.Dzhunushaliev, Izv. vuzov, ser. Fizika, N3, 108(1995) (In Russian).
\bibitem{wh}
C.Misner, K.Thorne,  J.Wheeler,  Gravitation  (W.H.Freeman
and Company, San Francisco, 1973).

\end{references}
\end{document}